\documentclass[12pt]{iopart}

\usepackage{iopams}  
\usepackage{graphicx}
\usepackage{color}

\newcommand{\text}[1]{\mathrm{#1}} 
\newcommand{\eqref}[1]{(\ref{#1})}
\newcommand{\Eqref}[1]{Eq.~\eqref{#1}}

\begin{document}

\title{Geothermal Casimir Phenomena}

\author{Klaus Klingm\"uller$^{1,2}$ and Holger Gies$^1$}

\address{$^1$ Institute for Theoretical Physics, Heidelberg
  University, Philosophenweg 16, D-69120 Heidelberg, Germany \\
$^2$ Institut f\"ur Theoretische Physik E, RWTH Aachen, D-52056 Aachen, Germany}
\ead{h.gies@thphys.uni-heidelberg.de, klingmueller@physik.rwth-aachen.de}
\begin{abstract}
  We present first worldline analytical and numerical results for the
  nontrivial interplay between geometry and temperature dependencies
  of the Casimir effect. We show that the temperature dependence of
  the Casimir force can be significantly larger for open geometries
  (e.g., perpendicular plates) than for closed geometries (e.g.,
  parallel plates). For surface separations in the experimentally
  relevant range, the thermal correction for the perpendicular-plates
  configuration exhibits a stronger parameter dependence and exceeds
  that for parallel plates by an order of magnitude at room
  temperature. This effect can be attributed to the fact that the
  fluctuation spectrum for closed geometries is gapped, inhibiting the
  thermal excitation of modes at low temperatures. By contrast, open
  geometries support a thermal excitation of the low-lying modes in
  the gapless spectrum already at low temperatures.
\end{abstract}


\section{Introduction}

The Casimir effect \cite{Casimir:dh} is a paradigm for
fluctuation-induced phenomena. Casimir forces between mesoscopic or
even macroscopic objects which result from fluctuations of the
ubiquitous radiation field or of the charge distribution on the
objects inspire many branches of physics, ranging from mathematical to
applied physics, see \cite{Bordag:2001qi} for reviews. Since
fluctuations usually occur on all momentum or length scales, they
encode both local as well as global properties of a given system. In
the case of the Casimir effect, the resulting force is influenced by
localized properties of the involved objects such as surface roughness
as well as by the global geometry of a given configuration. From a
technical perspective, localized properties can often be taken into
account by perturbative methods owing to a separation of scales: e.g.,
the corrugation wavelength and amplitude are usually much smaller than
the object's separation distance. But global properties such as
geometry or curvature dependencies generally require a full
understanding of the fluctuation spectrum in a given configuration.

Recent years have witnessed the development of a variety of new
field-theoretical methods for understanding and computing fluctuation
phenomena. So far, only phenomenological recipes had early been
developed for more complex Casimir geometries, such as
the proximity force approximation (PFA) \cite{pft1}. For the special
case of Casimir forces between compact objects, field-theoretic
results in asymptotic limits had been worked out
\cite{Feinberg,Balian:1977qr}. A first field-theoretic study of the
experimentally important configuration of a sphere above a plate
\cite{Lamoreaux:1996wh}
was performed in \cite{semicl} based on a semiclassical expansion. A
constrained functional-integral approach, as first introduced in
\cite{Bordag:1983zk} for the parallel-plate case, was further
developed for corrugated surfaces in \cite{Emig:2001dx}.

The sphere-plate as well as cylinder-plate configuration
\cite{Brown-Hayes:2005uf} was also used as a first example for the
worldline approach to the Casimir effect \cite{Gies:2003cv}, which is
based on a mapping of field-theoretic fluctuation averages onto
quantum-mechanical path integrals. This technique is rooted in the
string-inspired approach to quantum field theory which is particularly
powerful for the computation of amplitudes and effective actions in
background fields \cite{Schmidt:1993rk}. For arbitrary backgrounds,
the path integral over the worldlines representing the spacetime
trajectories of the quantum fluctuations can straightforwardly be
computed by Monte Carlo methods, as first demonstrated in
\cite{Gies:2001zp}. The particular advantage of the approach arises
from the fact that the computational algorithm can be formulated
independently of the background. This makes the approach so valuable
for Casimir problems, where a given surface geometry constitutes the
background for the fluctuations. The resulting technical
simplifications become particularly transparent for fluctuations
obeying Dirichlet boundary conditions (b.c.), where high-precision
computations have been performed, e.g., for the sphere-plate and
cylinder-plate case \cite{Gies:2005ym,Gies:2006bt,Gies:2006cq}.

A number of further first-principles approaches for arbitrary Casimir
geometries have been developed and successfully applied in recent
years. The constraint functional-integral approach has been extended
to general dispersive forces between deformed media
\cite{Emig:2003eq}. In particular, approaches based on scattering
theory have proved most successful, starting with an exact study of
the sphere-plate configuration with Dirichlet
b.c.~\cite{Bulgac:2005ku}. Scattering theory also lead to a solution
for the cylinder-plate case which, as a waveguide configuration,
allowed for a study of the case with real electromagnetic
b.c.~\cite{Emig:2006uh}. These scattering tools have been further
developed to facilitate an analytical computation of the important
small-curvature expansion \cite{Bordag:2006vc}.  For configurations
with compact objects, new scattering formulations have recently been
found which separate the problem into the scattering off the single
objects on the one hand and a propagation of the fluctuation
information between the objects on the other hand
\cite{Kenneth:2006vr,Emig:2007cf}; in particular, electromagnetic
b.c.~for real materials can conveniently be addressed with a new
formulation which emphasizes the charge fluctuations on the surfaces
\cite{Emig:2007cf}. Let us also mention the combination of scattering
theory with a perturbative expansion that has recently allowed to
study geometry effects beyond the PFA \cite{Rodrigues:2006ku}.
Scattering theory is also a valuable tool for analyzing Casimir
self-energies \cite{Graham:2003ib}. Finally, direct mode summation has
also successfully been applied to nontrivial geometries
\cite{Mazzitelli:2006ne}. 

In a real Casimir experiment, further properties such as finite
conductivity, surface roughness and finite temperature have to be
accounted for in addition to the geometry. Generically, these
corrections do not factorize but reveal a nontrivial interplay. For
instance, the interplay between dielectric material properties and
finite temperature \cite{Sernelius} still seems insufficiently
understood and has lead to a long-standing controversy
\cite{Mostepanenko:2005qh,Brevik:2006jw}. In the present work, we
confine ourselves to the ideal Casimir effect where this controversy
does not exists; but even in the ideal limit, the interplay between
geometry and temperature can be substantial, as demonstrated below. The
difference is not only of quantitative nature, but arises from the
underlying spectral properties of the fluctuations, as first pointed
out by Jaffe and Scardicchio \cite{Scardicchio:2005di}. In the
familiar parallel-plate case, the nontrivial part of the spectrum
transverse to the plates exhibits a gap of wave number
$k_{\mathrm{gap}}=\pi/a$, where $a$ is the plate separation. At
temperatures $T$ smaller than this gap, the relevant fluctuation modes
are hardly excited, implying a suppression of the thermal corrections;
the leading small-temperature contribution to the parallel-plates
Casimir force scales like $(aT)^4$. Geometries with a gap in the
relevant part of the excitation spectrum are called closed. Following
the same line of argument, we expect a suppression for thermal effects
for all closed geometries.

By contrast, there is no reason for this strong suppression of thermal
corrections in open geometries which do not have a gap in the
fluctuation spectrum. The sphere-plate or cylinder-plate cases belong
to this class. For open geometries, there are always Casimir-relevant
modes in the fluctuation spectrum that can be excited at any small
value of the temperature. Hence, we expect a much stronger dependence on
the temperature, e.g., $(aT)^\alpha$ with $0<\alpha<4$, and thus a
potentially much stronger thermal contribution in the experimentally
relevant parameter range $aT\sim 0.01\dots 0.1$.
 
So far, no first-principle calculation has been able to confirm this
expectation, since generic asymptotic-limit considerations and
standard approximations typically break down in the relevant parameter
range, as already emphasized in \cite{Scardicchio:2005di}. For
instance, the exact solution for the cylinder-plate case allowed for
an explicit temperature study of the limit of small cylinder radius, $R\ll
a,\beta$, where $\beta=1/T$. In this limit, a log-modified $(aT)^4$
correction is obtained for the dominant part of the spectrum with
Dirichlet b.c.~\cite{Emig:2006uh}. This result suggests that the
low-lying thermal excitations with long wavelength are not suppressed
by a gap but by the smallness of the cylinder radius required by the
asymptotic-limit considerations. 

Also, the use of recipes such as the PFA can lead to a different
scaling, such as a $(aT)^3$ law for the sphere-plate case
\cite{Lamoreaux:1996wh,Bordag:2000kf}. Whereas the PFA at zero
temperature is justifiable in the low-curvature limit, $a\ll R$,
\cite{semicl,Gies:2003cv,Scardicchio:2004fy,Gies:2006bt}, PFA-deduced
thermal corrections can be problematic: at small temperatures with thermal
wavelength much larger than the minimal surface separation $aT\ll 1$,
the thermal excitations can be more sensitive to the curvature radius
than the vacuum fluctuations. Even worse, the PFA uses the
parallel-plate formula, and hence a gapped spectrum, as an input and
thus misses the important difference arising from an open geometry.

\begin{figure*}[t]
	\centering
\includegraphics[width=0.35\linewidth]{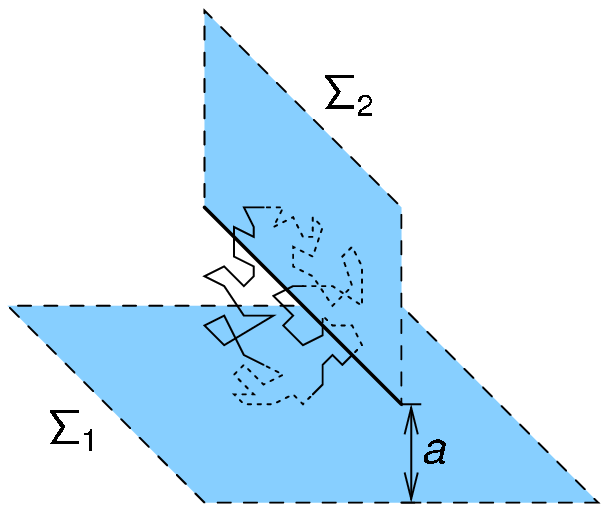}\hspace{0.5cm}
\includegraphics[width=0.55\linewidth,clip]{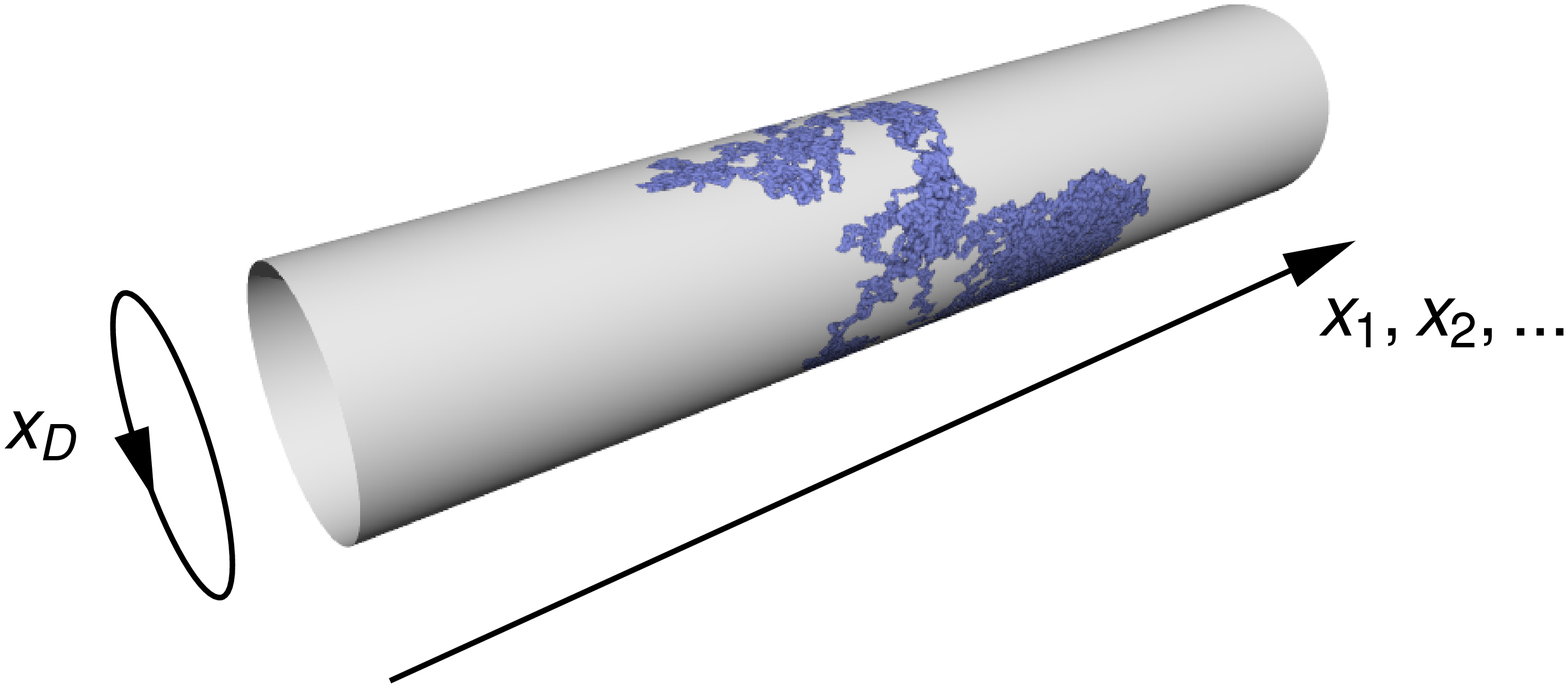}
\caption{Left panel: sketch of the parallel-plate configuration (taken
  from \cite{Gies:2006xe}). Right panel: sketch of the
  finite-temperature spacetime; a worldline can wind around the
  compactified time dimension.} 
\label{fig:sketches}
\end{figure*}

In this work, we present first evidence for a strong thermal
correction to a Casimir force law for an open geometry using worldline
numerics. As a paradigmatic example, we use the configuration of a
semi-infinite half-plate perpendicularly above an infinite plate
(cf. Fig.~\ref{fig:sketches}), imposing Dirichlet b.c.~for the
fluctuations of a real scalar field. This configuration belongs to a
set of cases, revealing a universal force law determined by
dimensionality, which has first been investigated in the context of
Casimir edge effects \cite{Gies:2006xe}. Since the configuration has
only one length scale which is the distance $a$ between the edge of
the half-plate and the infinite plate, the interplay of the gapless
fluctuation spectrum with thermal excitations is not
disturbed by other length scales, resulting in a clean thermal signature
of an open geometry. Our worldline calculations yield a thermal
correction obeying an $(aT)^3$ force law at low temperature. This implies a
substantial increase of the thermal contribution compared to those for
a closed geometry. 

The fact that geometry and temperature exhibit such a nontrivial
interplay in Casimir systems, resulting in ``geothermal'' Casimir
phenomena\footnote{We introduce the attribute ``geothermal'' here,
since it directly describes the source and nature of this
phenomenon. No link exists between the physics discussed here and, e.g.,
geothermal heat-pumps etc. dealt with in the geological sciences.}, is
another peculiar feature that should be added to the long list of
peculiarities of the Casimir effect; it clearly deserves further
investigation.

\section{Worldline approach to the Casimir effect at finite temperature}

Let us briefly summarize the worldline approach to the Casimir
effect. More detailed descriptions and derivations from first
principles can be found in \cite{Gies:2003cv,Gies:2006cq}. We consider
the Casimir interaction energy, serving as a potential energy for the
force, for two rigid objects with surfaces $\Sigma_1$ and
$\Sigma_2$. For a massless scalar field with Dirichlet boundaries in
$D=3+1$, the worldline representation of the Casimir interaction
energy is given by
\begin{equation}
E_{\mathrm{Casimir}}=-\frac{1}{2} \frac{1}{(4\pi)^2} \int_{0}^\infty
\frac{d \mathcal{T}}{\mathcal{T}^3}\int d^3x_\mathrm{CM} 
\, \left\langle\Theta_\Sigma[\mathbf x(\tau)]
\right\rangle . \label{eq:ECasW} 
\end{equation}
Here, the worldline functional $\Theta_\Sigma[\mathbf x(\tau)]=1$ if
the path $\mathbf x(\tau)$ intersects the surface
$\Sigma=\Sigma_1\cup\Sigma_2$ in both parts $\Sigma_1$ and $\Sigma_2$,
and $\Theta_\Sigma[\mathbf x(\tau)]=0$ otherwise%
.

This compact formula has an intuitive interpretation: the worldlines
can be viewed as the spacetime trajectories of the quantum
fluctuations of the scalar field. Any worldline that intersects the
surfaces does not satisfy Dirichlet boundary conditions. All
worldlines that intersect both surfaces thus should be removed from
the ensemble of allowed fluctuations, thereby contributing to the
negative Casimir interaction energy. The auxiliary integration
parameter $\mathcal{T}$, the so-called propertime, effectively governs
the extent of a worldline in spacetime. Large $\mathcal{T}$ correspond
to IR fluctuations with large worldlines, small $\mathcal{T}$ to UV
fluctuations.

The expectation value in \Eqref{eq:ECasW} has to be taken with respect
to the ensemble of worldlines that obeys a Gau\ss ian velocity
distribution
\begin{equation}
\langle \dots \rangle = 
{  \int_{\mathbf{x}_{\text{CM}}} \mathcal{D} x\,\dots\,  e^{-\frac{1}{4}
  \int_0^{\mathcal{T}} d\tau \, \dot{x}^2(\tau)}}\Bigg/
{
  \int_{\mathbf{x}_{\text{CM}}} \mathcal{D} x  e^{-\frac{1}{4}
  \int_0^{\mathcal{T}} d\tau \, \dot{x}^2(\tau)}},\label{eq:expval}
\end{equation}
where the worldlines have a common center of mass $x_{\text{CM}}$. At zero
temperature, the time component of the worldlines cancels out for static
objects, hence the straightforward Monte Carlo computation of
Eqs.~\eqref{eq:ECasW} and \eqref{eq:expval} can be restricted to the spatial
part.

Finite temperature can now easily be implemented with the aid of the Matsubara
formalism, and also the technical changes of the numerical algorithm are only
minor: The Euclidean time, say along the $D$th direction, is compactified to
the interval $[0,\beta]$ with periodic boundary conditions for bosonic
fluctuations. As a consequence, the worldlines can also wind around the
time dimension, see Fig.~\ref{fig:sketches}. It is convenient to write a given
loop $x(\tau)$ with winding number $n$ as sum of a loop with no winding,
$\tilde x(\tau)$, and a translation in time running from zero to $n \beta$
with constant speed,
\begin{equation}
	x_\mu(\tau)
	= \tilde x_\mu(\tau) + n\beta\frac{\tau}{\mathcal T} \delta_{\mu D}.
	\label{eq:winding}
\end{equation}
The path integral over the different winding number sectors labeled by
$n$ factorizes for static configurations, yielding
\begin{eqnarray}
	\fl\int_{x(0)=x(\mathcal T)}\mathcal Dx\ e^{-\int_0^{\mathcal T}d\tau \frac{\dot x^2}{4}}
	\ \cdots\ 
	&= \sum_{n=-\infty}^\infty e^{- \frac{n^2\beta^2}{4\mathcal T}}
		\int_{\tilde x(0)=\tilde x(\mathcal T)}\mathcal D\tilde x
			\ e^{-\int_0^{\mathcal T}d\tau \frac{\dot{\tilde x}^2}{4} }
			\ \cdots. \label{eq:windingsum}
\end{eqnarray}
The worldline representation of the Casimir interaction energy
for the Dirichlet scalar at finite temperature thus reads
\begin{equation}
\fl
E_\mathrm{Casimir}=-\frac{1}{2} \frac{1}{(4\pi)^2} \int_{0}^\infty
\frac{d \mathcal T}{\mathcal{T}^3}
\biggl( \sum_{n=-\infty}^\infty e^{- \frac{n^2\beta^2}{4\mathcal T}} \biggr)
\int d^3x_\mathrm{CM} 
\, \left\langle\Theta_\Sigma[ \mathbf x(\tau)]
\right\rangle . \label{eq:Etemp} 
\end{equation}
Whereas the worldline expectation value remains identical to the one
at zero temperature, the winding sum re-weights the propertime
integrand: larger temperature emphasizes smaller propertimes and vice
versa. This confirms the expectation that thermal corrections at low
temperature are dominated by long wavelength fluctuations which in our
case correspond to worldlines with a large spatial extent.

It is important to note that $E_{\text{Casimir}}$ is normalized such
that $E_{\text{Casimir}}\to 0$ for infinite distances $a\to \infty$.
Hence, $E_{\text{Casimir}}$ can differ from the thermodynamic free
energy by thermal corrections to the self-energies of the single
surfaces. The latter is distance-independent and thus does not
contribute to the Casimir force.

\subsection{Parallel plates}

As a test, let us consider the two parallel plates separated by a
distance $a$ along the $z$~axis.  Interchanging expectation
value and $z_\mathrm{CM}$ integration in Eq.~(\ref{eq:Etemp}), we encounter
\begin{equation}
	\int_{-\infty}^\infty\, dz_{\mathrm{CM}}\,
	\Theta_\Sigma[\mathbf x]
	=
	(\sqrt{\mathcal T} l - a)\, \theta (\sqrt{\mathcal T} l -a) ,
\end{equation}
where $l$ denotes the dimensionless extent of the given worldline in
$z$~direction measured in units of $\sqrt{\mathcal T}$; cf. \cite{Gies:2006cq}.
Differentiating Eq.~(\ref{eq:Etemp}) by $-\partial/\partial a$ yields
the Casimir force
\begin{equation}
	F_\mathrm{Casimir}
	=
	-\frac{1}{2} \frac{A}{(4\pi)^2} \, \frac{1}{a^4} \left\langle
	\int_{1/l^2}^\infty
	\frac{d \mathcal T}{\mathcal T^3}
	\biggl( \sum_{n=-\infty}^\infty e^{-
	\frac{n^2}{4\mathcal T}\, \frac{\beta^2}{a^2}} \biggr)
	\right\rangle,\label{eq:pp}
\end{equation}
where $A$ is the (infinite) area of the plates.
\begin{figure}
	\centering
	\includegraphics[width=0.7\columnwidth]{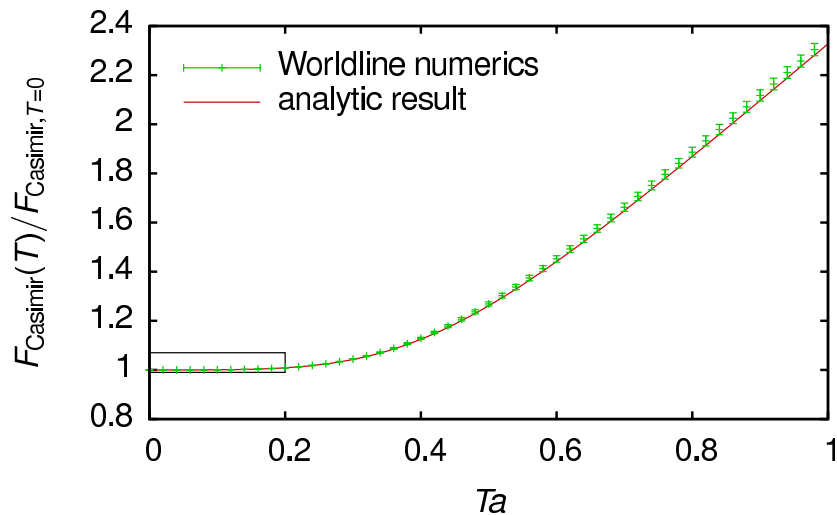}
	\caption{Temperature dependence of the Casimir force between two
	parallel plates. The ratio of the force
	at temperature $T$ and at
	zero temperature is plotted versus the temperature in units
	of the plate distance $a$. The rectangle at
	low temperature marks the region magnified in
	Fig.~\ref{fig:smalltemperature}.
	For the worldline numerical result, we have
	employed 800 loops each with $1\;000\;000$ points.}
	\label{fig:temperature}
\end{figure}
Figure \ref{fig:temperature} shows the numerical result for the Casimir
interaction energy \eqref{eq:pp}, corresponding to the 
distant-dependent part of the free energy, normalized to the zero-temperature
result. For comparison, the analytic result,
\begin{equation}
	\fl
	F_\mathrm{Casimir}
	=
	\frac{\pi^2}{2}\frac{\partial}{\partial a} \left[ \frac{AT}{a^2}
	\sum_{m=1}^{\infty} \frac{1}{(2 \pi m)^3} \left(
	\coth(2 \pi m a T)
	+ 2 \pi m a T \, \mathrm{csch}^2(2 \pi m a T)
	\right)\right],
\end{equation}
is also shown, see e.g. \cite{Feinberg:1999ys}\footnote{We use half
the value of \cite{Feinberg:1999ys} which is derived for the
electromagnetic field with two degrees of freedom.}. Both results
agree satisfactorily.

Incidentally, the leading thermal correction can be obtained
analytically from the worldline representation \eqref{eq:pp}: for
$(aT)^2 \ll 1$ (and $n\neq0$), the propertime integrand is dominated
by large $\mathcal T$, hence the lower bound can safely be set to
zero. This results in the well-known leading  thermal correction $\Delta F(T)=-
(\pi^2/90) A T^4$, which can also be understood as an excluded volume
effect: thermal modes are excluded from the region between the two
plates which thus does not contribute to the Stefan-Boltzmann law.

\subsection{Perpendicular plates}

We now consider the perpendicular-plate configuration introduced above.
Again, we can perform the $z_{\text{CM}}$ integration first, yielding
for the force 
\begin{equation}
	F_\mathrm{Casimir}
	=
	-\frac{L}{2 (4 \pi)^2}\, \frac{1}{a^3} \left\langle
	\int d\xi \int_{1/l(\xi)^2} \frac{d{\mathcal T}}{\mathcal{T}^{5/2}}
	\biggl( \sum_{n=-\infty}^\infty e^{-
	\frac{n^2}{4\mathcal T}\, \frac{\beta^2}{a^2}} \biggr) 
	\right\rangle,
	\label{eq:perpp}
\end{equation}
where $L$ is the (infinite) length of the system along the edge.
Here, $l(\xi)$ denotes the dimensionless extent of the given worldline
in $z$~direction as seen by the configuration in units of
$\sqrt{\mathcal T}$; it depends on the position $\xi$ of the worldline
normal to the perpendicular plate which is also measured in units of
$\sqrt{\mathcal T}$; for details, see  \cite{Klingmuller:2007}.
\begin{figure}
	\centering
	\includegraphics[width=0.7\columnwidth]{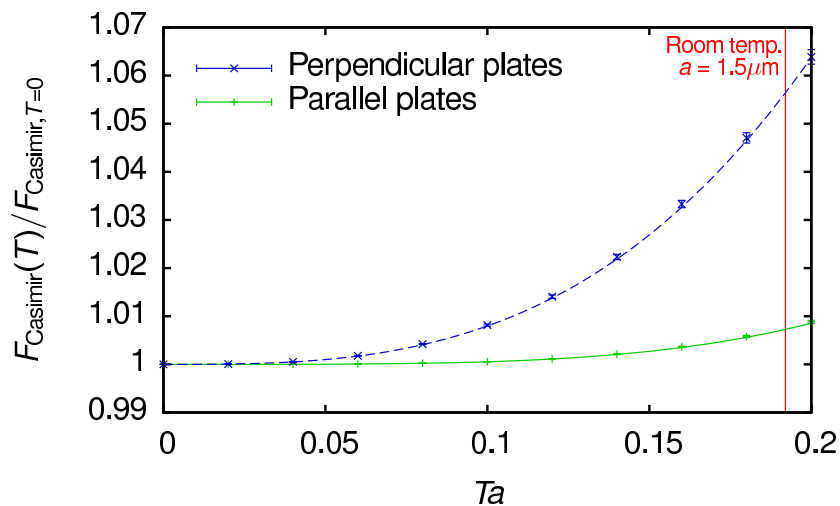}
	\caption{Temperature dependence of the Casimir force for two
	perpendicular plates compared to the parallel-plates result.
	The ratio of the force at temperature $T$ and at zero
	temperature is plotted versus the temperature in units of the
	plate distance $a$.  The plot shows the small-temperature
	range of Fig.~\ref{fig:temperature}. At experimentally
	relevant large-separation values of $T a$ (vertical line), the
	temperature correction for the perpendicular plates is $\simeq
	6 \%$, which should be compared with  $\sim0.7\%$ for the
	parallel plates. For the worldline numeric results we have
	employed 800 loops each with 1~000~000 points. The error bars
	represent the statistical error.}
	\label{fig:smalltemperature}
\end{figure}
Figure \ref{fig:smalltemperature} compares the resulting temperature
correction with that of the parallel-plates case in the small
temperature range of Fig.~\ref{fig:temperature}. In contrast to the
weak $(aT)^4$ dependence of the parallel-plates result, the Casimir
interaction energy for the perpendicular plates shows a strong increase with
temperature.  For typical experimental values at larger distance $a=1.5\mu$m and
room temperature, the temperature correction is about $6 \%$.  At the same
distance and temperature, the temperature effect for the parallel
plates is  $0.7 \%$. The open geometry therefore exhibits a
thermal correction which is  an order of magnitude larger
than the closed parallel-plates case. 

The leading thermal correction can again be computed analytically from
the worldline representation \eqref{eq:perpp} by extending the lower
bound of the $\mathcal T$ integral to 0 and using $\langle \int
d\xi\rangle \equiv \langle l\rangle=\sqrt{\pi}$.
Here, $l$ denotes the extension of the loop perpendicular to the
semi-infinite plate \cite{Gies:2006cq}. We obtain
\begin{equation}
F_{\text{Casimir}}(T)\simeq F_{\text{Casimir},{T=0}} -
\frac{\zeta(3)}{4\pi} \, \frac{L}{a^3}\, (a T)^3,\quad \text{for}\,\, (aT)\ll 1, 
\end{equation}
which is confirmed by the full numerical result over the whole range
of temperatures shown in Fig.~\ref{fig:smalltemperature}.  We note
that the low-temperature scaling of the thermal correction for this
open geometry cannot be understood as an excluded-volume effect.

\section{Conclusions}

We have presented analytical as well as numerical results for the
nontrivial interplay between geometry and finite temperature for the
Casimir effect in open geometries. For the first time, we have shown
that the gapless nature of the fluctuation spectrum leads to a strong
enhancement of the thermal correction to the Casimir force. Our
numerical data for the perpendicular-plate case with Dirichlet
b.c.~confirms our analytically derived $(aT)^3$ force law at low
temperatures. This should be compared to the weaker $(aT)^4$ dependence
of the parallel-plates case where a gap in the spectrum suppresses the
thermal correction. This calls urgently for further first-principles
computations for other open geometries such as the sphere-plate case.

\medskip

It is a pleasure to thank Michael Bordag and his team for the
organization of the QFEXT07 workshop and for creating such a stimulating
atmosphere. This work was supported by the DFG under Gi 328/1-3 (Emmy-Noether
program) and Gi 328/3-2.


\section*{References}
\begin{thebibliography}{10}


\bibitem{Casimir:dh}
H.B.G.~Casimir,
Kon.\ Ned.\ Akad.\ Wetensch.\ Proc.\  {\bf 51}, 793 (1948).

\bibitem{Bordag:2001qi}
  K.~A.~Milton,
{\it  River Edge, USA: World Scientific (2001)}; 
M.~Bordag, U.~Mohideen and V.~M.~Mostepanenko,
Phys.\ Rept.\  {\bf 353}, 1 (2001); 
%
  S.~Y.~Buhmann and D.~G.~Welsch,
  Prog.\ Quant.\ Electron.\  {\bf 31}, 51 (2007)
  [arXiv:quant-ph/0608118].

\bibitem{pft1}
B.V.~Derjaguin, I.I.~Abrikosova, E.M.~Lifshitz, Q.Rev. {\bf 10}, 295
(1956); 
J.~Blocki, J.~Randrup, W.J.~Swiatecki, C.F.~Tsang, Ann.~Phys.~(N.Y.)
{\bf 105}, 427 (1977).
 
\bibitem{Feinberg}
G.~Feinberg and J.~Sucher,
Phys.\ Rev.\ A {\bf 2}, 2395 (1970).

\bibitem{Balian:1977qr}
  R.~Balian and B.~Duplantier,
  Annals Phys.\  {\bf 112}, 165 (1978).

\bibitem{Lamoreaux:1996wh}
S.~K.~Lamoreaux,
Phys.\ Rev.\ Lett.\  {\bf 78}, 5 (1997); 
%
U.~Mohideen and A.~Roy,
Phys.\ Rev.\ Lett.\  {\bf 81}, 4549 (1998);
%
H.B.~Chan {\em et al.}, 
Science 291, 1941 (2001); 
%
%
R.S.~Decca {\em et al.}, 
Phys.\ Rev.\ D {\bf 68}, 116003 (2003); 
Phys.\ Rev.\ Lett.\  {\bf 94}, 240401 (2005).

\bibitem{semicl}
M. Schaden and L. Spruch, Phys.~Rev.~A {\bf 58}, 935 (1998);
Phys. Rev. Lett. {\bf 84} 459 (2000) 

\bibitem{Bordag:1983zk}
  M.~Bordag, D.~Robaschik and E.~Wieczorek,
  Annals Phys.\  {\bf 165}, 192 (1985).

\bibitem{Emig:2001dx}
T.~Emig, A.~Hanke and M.~Kardar,
Phys.\ Rev.\ Lett.\  {\bf 87} (2001) 260402.

\bibitem{Brown-Hayes:2005uf}
M.~Brown-Hayes, D.A.R.~Dalvit, F.D.~Mazzitelli, W.J. Kim and R.~Onofrio,
Phys.\ Rev.\ A {\bf 72}, 052102 (2005).

\bibitem{Gies:2003cv}
H.~Gies, K.~Langfeld and L.~Moyaerts,
JHEP {\bf 0306}, 018 (2003); 
arXiv:hep-th/0311168.
 
\bibitem{Schmidt:1993rk}
  M.~G.~Schmidt and C.~Schubert,
  Phys.\ Lett.\  B {\bf 318}, 438 (1993)
  [arXiv:hep-th/9309055]; 
for a review, see C.~Schubert,
Phys.\ Rept.\  {\bf 355}, 73 (2001).

\bibitem{Gies:2001zp}
H.~Gies and K.~Langfeld,
Nucl.\ Phys.\ B {\bf 613}, 353 (2001); 
Int.\ J.\ Mod.\ Phys.\ A {\bf 17}, 966 (2002).

\bibitem{Gies:2005ym}
H.~Gies and K.~Klingmuller,
 J.~Phys.~A {\bf 39} 6415 (2006) [arXiv:hep-th/0511092].

\bibitem{Gies:2006bt}
H.~Gies and K.~Klingmuller,
Phys.\ Rev.\ Lett.\  {\bf 96}, 220401 (2006)
[arXiv:quant-ph/0601094].

\bibitem{Gies:2006cq}
  H.~Gies and K.~Klingmuller,
  Phys.\ Rev.\  D {\bf 74}, 045002 (2006)
  [arXiv:quant-ph/0605141].
 
\bibitem{Emig:2003eq}
T.~Emig and R.~Buscher,
Nucl.\ Phys.\ B {\bf 696}, 468 (2004).

\bibitem{Bulgac:2005ku}
A.~Bulgac, P.~Magierski and A.~Wirzba,
Phys.\ Rev.\ D {\bf 73}, 025007 (2006)
[arXiv:hep-th/0511056];
A.~Wirzba, A.~Bulgac and P.~Magierski,
J.\ Phys.\ A {\bf 39} (2006) 6815
[arXiv:quant-ph/0511057].

\bibitem{Emig:2006uh}
T.~Emig, R.~L.~Jaffe, M.~Kardar and A.~Scardicchio,
Phys.\ Rev.\ Lett.\  {\bf 96} (2006) 080403.

\bibitem{Bordag:2006vc}
  M.~Bordag,
  Phys.\ Rev.\  D {\bf 73}, 125018 (2006); 
  Phys.\ Rev.\  D {\bf 75}, 065003 (2007).

\bibitem{Kenneth:2006vr}
  O.~Kenneth and I.~Klich,
  Phys.\ Rev.\ Lett.\  {\bf 97}, 160401 (2006); 
  arXiv:0707.4017.
%

\bibitem{Emig:2007cf}
  T.~Emig, N.~Graham, R.~L.~Jaffe and M.~Kardar,
  arXiv:0707.1862; 
  arXiv:0710.3084. 

\bibitem{Rodrigues:2006ku}
  R.~B.~Rodrigues, P.~A.~Maia Neto, A.~Lambrecht and S.~Reynaud,
  Phys.\ Rev.\ Lett.\  {\bf 96}, 100402 (2006)
  [arXiv:quant-ph/0603120]; 
  Phys.\ Rev.\  A {\bf 75}, 062108 (2007).

\bibitem{Graham:2003ib}
N.~Graham, R.~L.~Jaffe, V.~Khemani, M.~Quandt, O.~Schroeder and H.~Weigel,
Nucl.\ Phys.\ B {\bf 677}, 379 (2004)
[arXiv:hep-th/0309130]; 
%

\bibitem{Mazzitelli:2006ne}
  F.~D.~Mazzitelli, D.~A.~R.~Dalvit and F.~C.~Lombardo,
  New J.\ Phys.\  {\bf 8}, 240 (2006); 
  D.~A.~R.~Dalvit, F.~C.~Lombardo, F.~D.~Mazzitelli and R.~Onofrio,
  Phys.\ Rev.\  A {\bf 74}, 020101 (2006).

\bibitem{Sernelius}
M.~Bostr\"om and Bo E.~Sernelius, Phys.~Rev.~Lett.~{\bf 84}, 4757 (2000).

\bibitem{Mostepanenko:2005qh}
V.~M.~Mostepanenko {\it et al.},
J.\ Phys.\ A {\bf 39}, 6589 (2006)
[arXiv:quant-ph/0512134].

\bibitem{Brevik:2006jw}
I.~Brevik, S.~A.~Ellingsen and K.~A.~Milton,
arXiv:quant-ph/0605005.

\bibitem{Scardicchio:2005di}
A.~Scardicchio and R.~L.~Jaffe,
Nucl.\ Phys.\ B {\bf 743} (2006) 249
[arXiv:quant-ph/0507042].

\bibitem{Bordag:2000kf}
  M.~Bordag, B.~Geyer, G.~L.~Klimchitskaya and V.~M.~Mostepanenko,
  Phys.\ Rev.\ Lett.\  {\bf 85}, 503 (2000).

\bibitem{Scardicchio:2004fy}
A.~Scardicchio and R.~L.~Jaffe,
Nucl.\ Phys.\ B {\bf 704}, 552 (2005);
Phys.\ Rev.\ Lett.\  {\bf 92}, 070402 (2004).

\bibitem{Gies:2006xe}
  H.~Gies and K.~Klingmuller,
  Phys.\ Rev.\ Lett.\  {\bf 97}, 220405 (2006)
  [arXiv:quant-ph/0606235].


\bibitem{Feinberg:1999ys}
  J.~Feinberg, A.~Mann and M.~Revzen,
  Annals Phys.\  {\bf 288} (2001) 103
  [arXiv:hep-th/9908149].

\bibitem{Klingmuller:2007}
K.~Klingmuller, Dissertation, Heidelberg U.~(2007).
 
\end{thebibliography}
\end{document}